\documentclass[amsfonts,notitlepage,aps,prl,twocolumn,11pt]{revtex4-1}
\usepackage{amsmath,amssymb,color,graphicx,mathrsfs,titlesec,hyperref,enumerate,bm}

\newcommand{\FF}{\mathscr{F} }

\definecolor{linkColor}{rgb}{1,0,0}
\hypersetup{pdfborder={0 0 0},colorlinks=true,urlcolor=linkColor,citecolor=linkColor}

\begin{document}

\title{Correcting nonlinear drift distortion of scanning probe microscopy from image pairs with orthogonal scan directions}
\author{Colin Ophus}
\affiliation{National Center for Electron Microscopy, Molecular Foundry, Lawrence Berkeley National Laboratory, Berkeley, CA}
\author{Chris Nelson}
\affiliation{Materials Science and Engineering, University of California Berkeley, Berkeley, CA}

\author{Jim Ciston}
\affiliation{National Center for Electron Microscopy, Molecular Foundry, Lawrence Berkeley National Laboratory, Berkeley, CA}

\begin{abstract}

Unwanted motion of the probe with respect to the sample is a ubiquitous problem in scanning probe microscopy, causing both linear and nonlinear artifacts in experimental images.  We have designed a procedure to correct these artifacts by using orthogonal scan pairs to align each measurement line-by-line along the slow scan direction.  We demonstrate the accuracy of our algorithm on both synthetic and experimental data and provide an implementation of our method.

\end{abstract}

\maketitle

\section{Introduction \label{intro}}


Scanning probe microscopy (SPM) is a very powerful experimental tool due in part to the small probe size and corresponding high spatial resolution of contemporary physical and focused radiation probes.  SPM experiments, such as scanning transmission electron microscopy (STEM), requires that the probe be moved across the sample surface in order to form an image.  This procedure can introduce artifacts in the measurement due to the time delay between measurements and the accumulation of error in the probe position, for example from drift of the sample \cite{harrach1995instrumental, borisevich2010mapping, berkels2012high, jones2013identifying, berkels2014optimized, sang2014revolving, zuo2014lattice, yankovich2014picometre}.  Examples of these artifacts include linear distortions such as shear, expansion or contraction applied to the whole image, random ``jitter'' of each scanline's origin position with respect to the intended position on the sample, jump discontinuities due to large sample jumps, and others.  Virtually all scanning probe experiments contain image distortions, and these distortions are often large compared to atomic-scale features.


Previous studies by other groups have attempted to measure and correct SPM distortions.    Berkels et al. recorded and aligned series of STEM image exposures to improve peak precision \cite{berkels2012high}, and have also devised a method for nonlinear registration of images series \cite{berkels2014optimized}.  Jones and Nellist corrected both linear and some nonlinear distortions in single STEM images by assuming prior knowledge of atomic features \cite{jones2013identifying}.  Sang and LeBeau have developed the ``REV-STEM'' method, where linear drift coefficients are measured and corrected by the a series of STEM images recorded in different scan directions \cite{sang2014revolving}.  In STEM experiments, many samples will be modified or damaged by excessive electron dose \cite{egerton2004radiation, Egerton2013} and therefore using as few measurements as possible to correct drift distortions is desirable.  To that end, we have created a general case algorithm which requires no \emph{a priori} assumptions about image features and requires only two images as an input, minimizing acquisition time and electron dose needed for correction of drift artifacts.


In this manuscript, we first show how nonlinear drift can affect scanning probe images.  We then develop an algorithm to correct all linear and nonlinear drift distortions in scanning probe images by correcting the scanline origin positions from two or more SPM images.   We generate corrected images by using kernel density estimation to resample the images, and a develop a Fourier weighting scheme to further reduce error.  We test this algorithm on synthetic data, and both simple and complex experimental datasets.  We evaluate the algorithm by measuring deviations of atomic sites in corrected images from the best-fit lattice positions, and by measuring complex lattice strain fields before and after correction.


\section{Theory}
\subsection{The directionality of scanning probe microscopy}


In a standard scanning probe experiment, data is recorded pixel-by-pixel sequentially.  The resulting array consists of a ``fast'' direction, consisting of the primary travel direction of the probe, and the ``slow'' direction, along which the probe is moved after each line is completed.  Because adjacent pixels in the fast scan direction are measured much closer together in time, the information transfer along this direction is more reliable than the slow direction.  After each line is completed, the probe must be quickly moved over a long distance, and repositioned directly below the preceding line, along the slow direction.  Information transfer along this direction is less reliable, because it is recorded much more slowly, and depends on the ability of the probe to be positioned with perfect accuracy after a change in probe speed and direction.  If the sample or the electronic set-points of the microscope drift during the experiment, this motion will affect the slow direction much more strongly than the fast direction.

In this paper, in order to develop an algorithm to correct scan distortions, we make the following assumptions about the scanning probe experiments:
\begin{enumerate}
  \item The scan directions are accurately known in the microscope frame of reference.
  \item The probe translation steps along the fast scan directions have negligible error. 
  \item The sample is unchanged during or between different scans.
\end{enumerate}
Given these assumptions, we will show that orthogonal scans can be used to correct both linear and nonlinear scan distortions.


\subsection{The effects of linear and nonlinear drift distortion on scanning probe microscopy \label{SectionDistortion}}

\begin{figure}[htbp]
    \centering
        \includegraphics[width=3.00in]{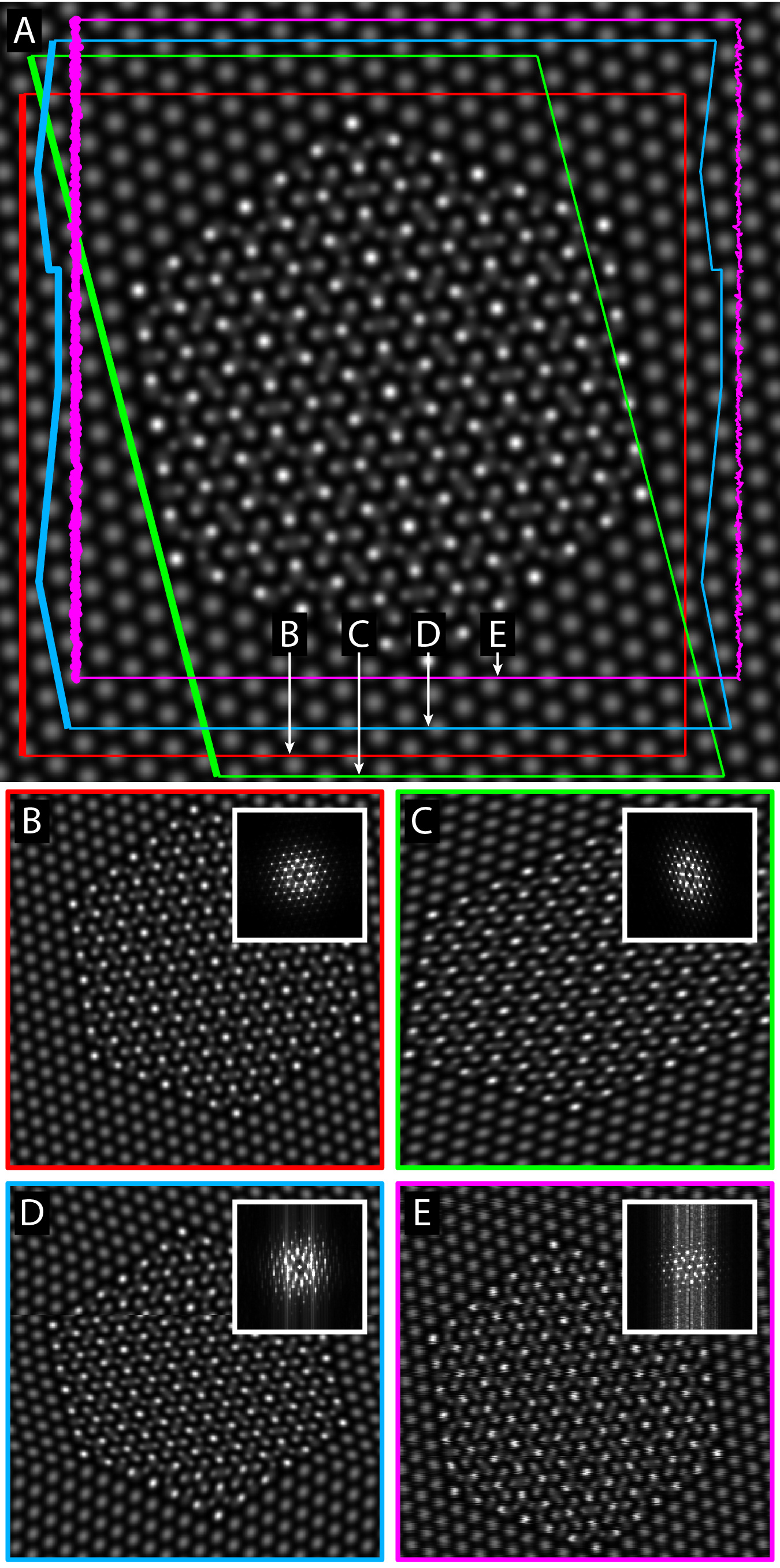}
 	\caption{(A) Synthetic dataset with examples of scanning probe microscopy images recorded with and without error in the probe positions, depicted by the colored outline where the scanline origins are represented by a thicker line.  The cases considered are (B) perfect sampling, (C) linear drift distortion, (D) nonlinear drift distortions and a jump discontinuity, and (E) random noise added to the scanline origins.  The square root of the Fourier transform amplitude is inset into each image.
		\label{FigureSyntheticExamples}}
\end{figure}

Fig.\ \ref{FigureSyntheticExamples}A shows a synthetic image consisting of two different hexagonal lattices with an epitaxial relationship.  One lattice represents a simple matrix, and the other forms a moire pattern with the matrix, inside a simulated precipitate, forming an image with both low and high spatial frequencies.  In an ideal scanning probe experiment, the sample remains perfectly stationary during collection of the data, and the probe is perfectly positioned with respect to the sample.  An ideal resampled scanning probe image is shown in Fig.\ \ref{FigureSyntheticExamples}B, with a horizontal fast scan direction.  This image is a perfect representation of the input data, since the sampling rate is below the Nyquist limit.

Scan distortions along the slow scan direction can be equivalently represented by either moving the sample with respect to an ideal scan, or by moving the scanline origins with respect to the sample.  Moving the sample continuously during a scanning probe experiment produces a linear drift artifact, like that shown in Fig.\ \ref{FigureSyntheticExamples}C.  This drift artifact can be described by a linear affine transformation of the underlying data \cite{bracewell1993affine}, consisting of an expansion in one direction and a compression in another.  Drift of this type is routinely corrected in scanning probe experiments \cite{horcas2007wsxm, rahe2010vertical, sang2014revolving}.

Examples of more complex drift artifacts are shown in Fig.\ \ref{FigureSyntheticExamples}D.  The artifacts could be caused by sudden jumps in the sample position, jumps in the probe positioning system, charging of the sample over time, or changes in the sample-probe environment.  In this example, the drift direction changes several times during data collection.  Additionally, a jump discontinuity is present roughly one third from the top of the image.  The simulated SPM image shows large distortions, and the inset Fourier transform shows streaks along the slow scan direction.  The streaks are caused by the various locally-distorted regions of the sample being misaligned with respect to each other.

Between scanlines, the probe must be repositioned directly adjacent to the origin of the previous scanline with high precision.  If there is some error (perhaps due to electronic noise or hysteresis), it could manifest as a random deviation between the ideal and the actual scanline origins.  An example of this is shown in Fig.\ \ref{FigureSyntheticExamples}E, which is very similar to flagging often observed in scanning probe microscopies.  This artifact also produces strong streaks along the slow scan direction in the Fourier transform.  The combination of the above artifacts can represent many nonlinear drift distortions observed in experiments.  In the next sections, we will show that these distortions can be corrected with high accuracy.

\subsection{An algorithm to correct nonlinear distortions using orthogonal scan directions}

\begin{figure}[htbp]
    \centering
        \includegraphics[width=3.00in]{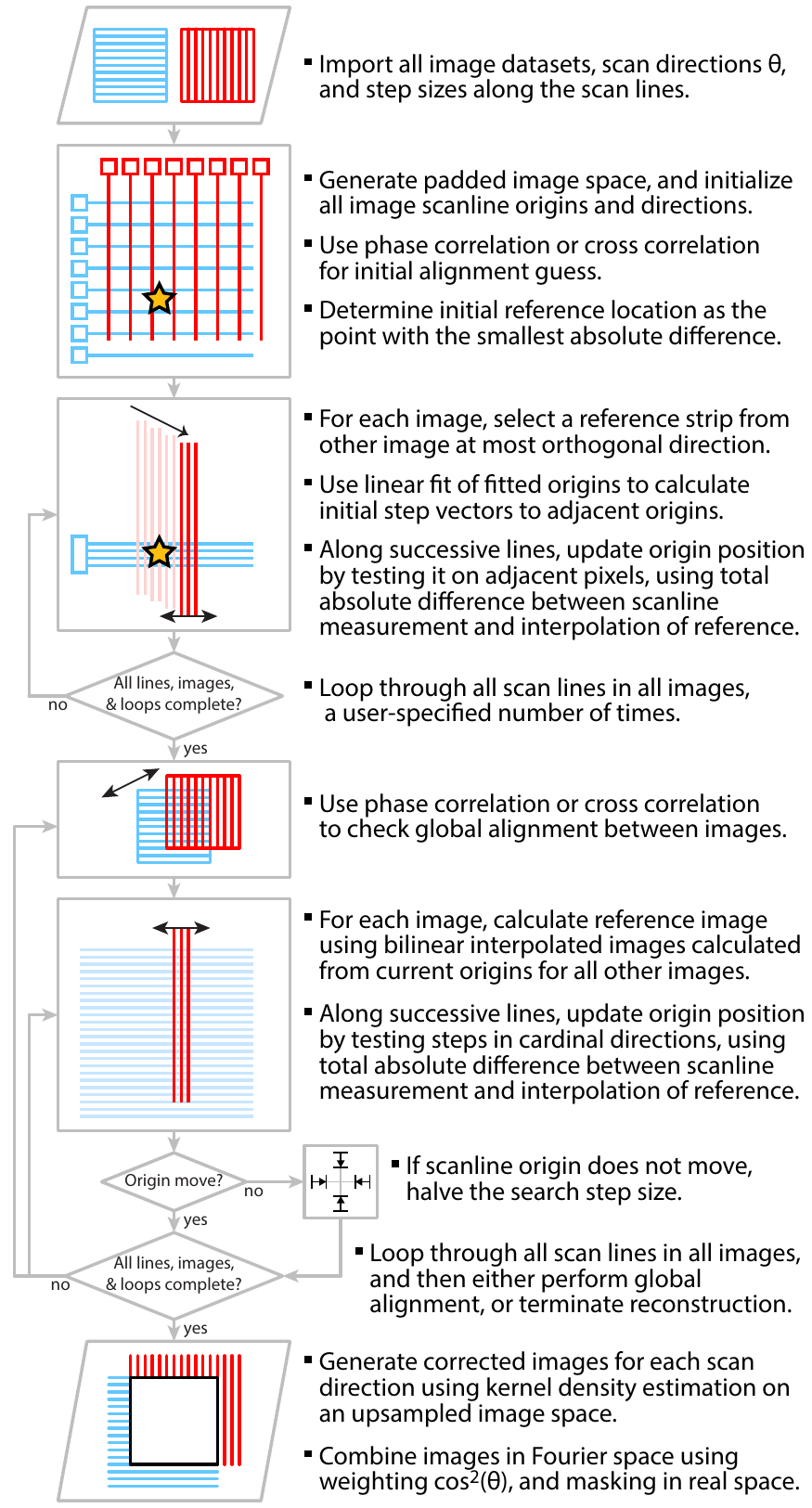}
 	\caption{Schematic of the nonlinear correction algorithm.}
	\label{FigureFlowChart}
\end{figure}

As in the previous section, we will represent all image distortions by translation of the scanline origins.  In order to reverse these distortions, we therefore need to estimate all scanline origins from the distorted data.  From a single scan, this is impossible without making some assumptions about the underlying data (such as assuming a periodic lattice).  However, we can make use of the accurate information transfer along the fast scan direction, which we assume to be essentially error-free.  By recording multiple images along orthogonal directions, we can use the fast scan direction of one image to calibrate the slow scan direction of another.  This procedure is outlined in Fig.\ \ref{FigureFlowChart}.

This algorithm requires a minimum of two scanning probe images, ideally recorded with fast scan directions 90 degrees apart.  We begin by initializing a reconstruction space that is large enough to contain all transformed and translated images.  Samples with a large amount of drift will require a larger padding region.  The images are first rotated to the specified scan directions, and roughly aligned using phase correlation.  After applying a low pass filter, a reference point is selected where the mean absolute difference between all images reaches a minimum. 

The next step is a rough initial alignment of all images.  Each image is aligned relative to a strip of user-defined width taken from whichever other image was recorded at a direction closest to 90 degrees away, starting from the reference point and moving outwards.  Each scanline is scored by taking the mean absolute difference between the measured intensity and a resampled estimate of the reference strip.  Next the scanline origin is moved to the four adjacent pixels and the associated score is computed.  Whichever scanline origin produces the lowest score is selected, and the origin position is updated.  This update is stabilized by under-relaxation, i.e. rather than moving one pixel in a given direction, the scanline will be moved in that direction by some reduced factor.  As the alignment proceeds outwards from the reference point, new scanline origins are generated by linearly fitting the step size between all fitted origins, and adding this step to adjacent origins sequentially.  The purpose of this rough alignment is to robustly estimate the global linear drift between different images and provide a good starting guess for the next step.  These iterations are repeated several times and can be checked visually by the user at this stage.

After the initial alignment is complete, we iteratively perform the main alignment of all datasets.  First, a synthetic image is generated for each recorded dataset using kernel density estimation (KDE), which is described in the Appendix. \ref{SectionAppendixKDE}  After a synthetic image has been generated from all datasets, the primary refinement proceeds.  Each dataset is aligned to the average of all other datasets (at least one, but could be any number), one scanline at a time.  This refinement is similar to that described above, where each scanline origin is scored using mean absolute difference and compared to an origin shift in each of the 4 nearest-neighbor pixel directions, with the shift size typically initialized to 0.5 pixels.  However in this refinement step, if a given scanline origin does not move, its search step size is halved.  After all scanlines have been refined once, the synthetic images are recalculated.  At this stage, global alignment between all images can be checked via phase correlation.  If an image shift of more than a few pixels is detected, the origin shift step size is reset to the initial value for that image.   These steps are repeated until either the total origin shift falls below a convergence threshold (typically 1 pixel per image), or a user-specified number of iterations are completed ($\approx$20 iterations required for a typical experimental STEM dataset).

At this stage the user again reviews the transformed images, and if required performs more primary refinement iterations.  Once the refinement is completed, the final step is to generate an output image.  This can be done simply by taking the mean of the transformed images.  However, this method is not ideal; the nonlinear drift correction algorithm described above cannot fully reconstruct the original data since some information can be irrevocably lost along the slow scan direction, due to noise or large nonlinear drift distortions.  A better method is to weight the fast scan direction of each transformed image more strongly than its slow scan direction.  This is accomplished by taking the Fourier transform of each image $I_i$, and multiplying by the weighting function $\cos^2 \theta$ where $\theta$ is the angle between each Fourier coordinate pixel and the scan direction.  The $N$ weighted images are then added together, and divided by the sum of all weighting functions.  The final image $I_{\rm{output}}$ is given by the inverse Fourier transform of the weighted sum:
\begin{equation}
	\FF \{ I_{\rm{output}} \} = \frac{ \sum_i^N  \cos^2 \theta_i \FF \{ I_i \} }{  \sum_i^N  \cos^2 \theta_i   }
	\label{EquationFourierWeight}
\end{equation}
where $\FF \{  \}$ represents the 2D Fourier transform.  This weighting can be tightened further towards the fast scan direction if more scan angles are recorded between 0 and 90 degrees.  This weighting can improve the corrected image quality by suppressing noise in the slow scan direction.



Our method has several limitations:  performing drift correction line-by-line requires that the data contains enough variance along each line to be accurately aligned.  If the data does not contain sufficient high spatial frequency information, additional constraints must be applied.  One potential constraint would be to apply a moving average to the scanline origins, or to force the origins to follow a linear path.  We therefore recommend images at atomic resolution be recorded at a slight angle with respect to crystallographic planes.  Another potential limitation occurs when applying our method to nearly-perfect periodic lattices, such as atomic-resolution images aligned to a low-index crystallographic zone axis.  A global lattice misalignment between images can easily occur, and therefore we require enough low spatial frequency information (such as an edge, a precipitate or alignment markers) to detect and correct lattice misalignments.   High-dose STEM experiments can also erode or damage the sample being imaged \cite{egerton2004radiation, Egerton2013}.  Therefore the electron dose should be sufficiently low for the sample to remain unchanged between scans.  Finally, we have assumed near-perfect information transfer along the fast scan direction of each scanline.  Systematic errors along this direction, for example electronic noise at 60 Hz in the STEM scanning coils, will cause misalignment of the relative images. 


\subsection{Implementation of our algorithm in MATLAB}

We have provided an implementation of our algorithm programmed in MATLAB in the supplementary materials of this manuscript.  This implementation should be general enough to correct any experimental datasets, and contains many user-selectable settings to aid in reconstructions.  The code has been tested on datasets consisting of up to 500 images.  It has also been tested for simultaneously recorded images, for example bright field and dark field pairs in STEM.  A synthetic nonlinear drift example like that described below is also included.

\section{Results and Discussion}
\subsection{Synthetic dataset example}


\begin{figure*}[htbp]
    \centering
        \includegraphics[width=6.0in]{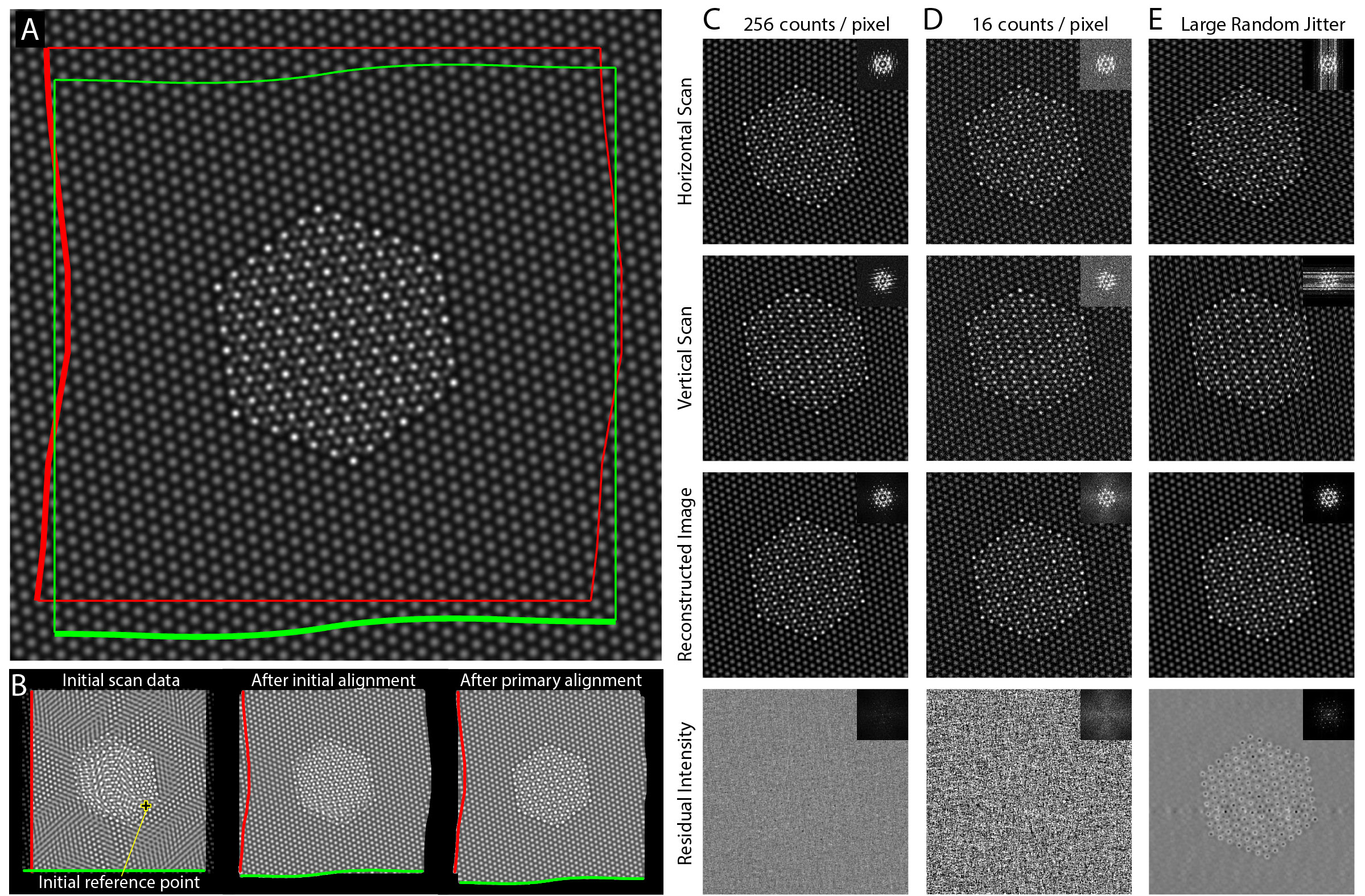}
 	\caption{(A) Synthetic image with scanline origins containing nonlinear drift shown as a colored outline, for a horizontal scan direction (red) and a vertical scan (green). (B) The three stages of drift correction applied to the example in (A). Drift correction examples for the data in (A) are shown with (C) high S/N (D) low S/N, and (E) scans with random Gaussian shifts (2 pixel standard deviation) of each scanline origin (``jitter'') added to the scanline origins shown in (A).  The first three rows of (C)-(E) show original and reconstructed images, bottom row shows residual noise after subtracting original signal with a range of $\pm 25 \%$ normalized intensity.  Windowed Fourier transforms inset into all images.}
	\label{FigureResultsSynthetic}
\end{figure*}

We first test the above algorithm using simulated image pairs.  Simulated tests are important because unlike in a real experiment, we know all of the original data values and can therefore precisely assess any residual error after correction.  The synthetic image data is plotted in Fig.\ \ref{FigureResultsSynthetic}A, with the scanline origins for an image pair with a large amount of nonlinear drift overlaid.  The progression of our drift correction algorithm applied to this dataset is shown in Fig.\ \ref{FigureResultsSynthetic}B.  After the initial alignment step to estimate image translation, several different moire lattice patterns are visible on the left image of Fig.\ \ref{FigureResultsSynthetic}B.  The rough alignment step removes the long-range nonlinear distortions, producing the smooth outer lattice shown in the center of Fig.\ \ref{FigureResultsSynthetic}B, though with a unit cell jump between the two images (visible in the precipitate in the center).  Finally, the main alignment step corrects the unit cell jump and removes all residual drift.  This can be seen by comparing the scanline origins in Fig.\ \ref{FigureResultsSynthetic}B to those of the original samples in Fig.\ \ref{FigureResultsSynthetic}A.

Two outputs of our nonlinear drift correction algorithm are shown in Figs.\ \ref{FigureResultsSynthetic}C-D, corresponding to high signal-to-noise (S/N) images, and low S/N images respectively. In both cases, the reconstructed image is very accurate and contains no obvious nonlinear drift artifacts.  The high S/N intensity residual is uniformly random and contains almost no residual signal of the crystal lattices.  The low S/N reconstruction has a similarly uniformly random residual.  However, the inset Fourier transform in Fig.\ \ref{FigureResultsSynthetic}D shows that the noise signal is biased along more horizontal and vertical directions.  This is because the noise is more correlated along the image pair scan directions, meaning that any error in the corrected scanline position will couple along the entire scanline.  Despite this issue, the reconstructed image is still a very good match to the original data, shown by the lack of residual crystalline peaks in the residual Fourier transform.

The drift correction example shown in Fig.\ \ref{FigureResultsSynthetic}E has infinite S/N, but contains random shifts of all scanline origins, with a magnitude approximately half the atomic spacing.  The reconstructed image closely matches the original synthetic data.  However, the residual image shows that the signal contrast (peak-to-valley ratio) is decreased, relative to the original data.  This is caused by the interpolation step of the reconstruction algorithm applied to the highly nonlinear sampling of the synthetic data.  Interpolation tends to reduce the deviation of an image from the local mean, leading to a residual signal in Fig.\ \ref{FigureResultsSynthetic}E approximately equal to the intensity mean minus the original data multiplied by a small negative value.  Despite this, the reconstructed image is still a good match to the original data, with the peak locations unchanged.

\subsection{Simple experimental example}


\begin{figure*}[htbp]
    \centering
        \includegraphics[width=6.0in]{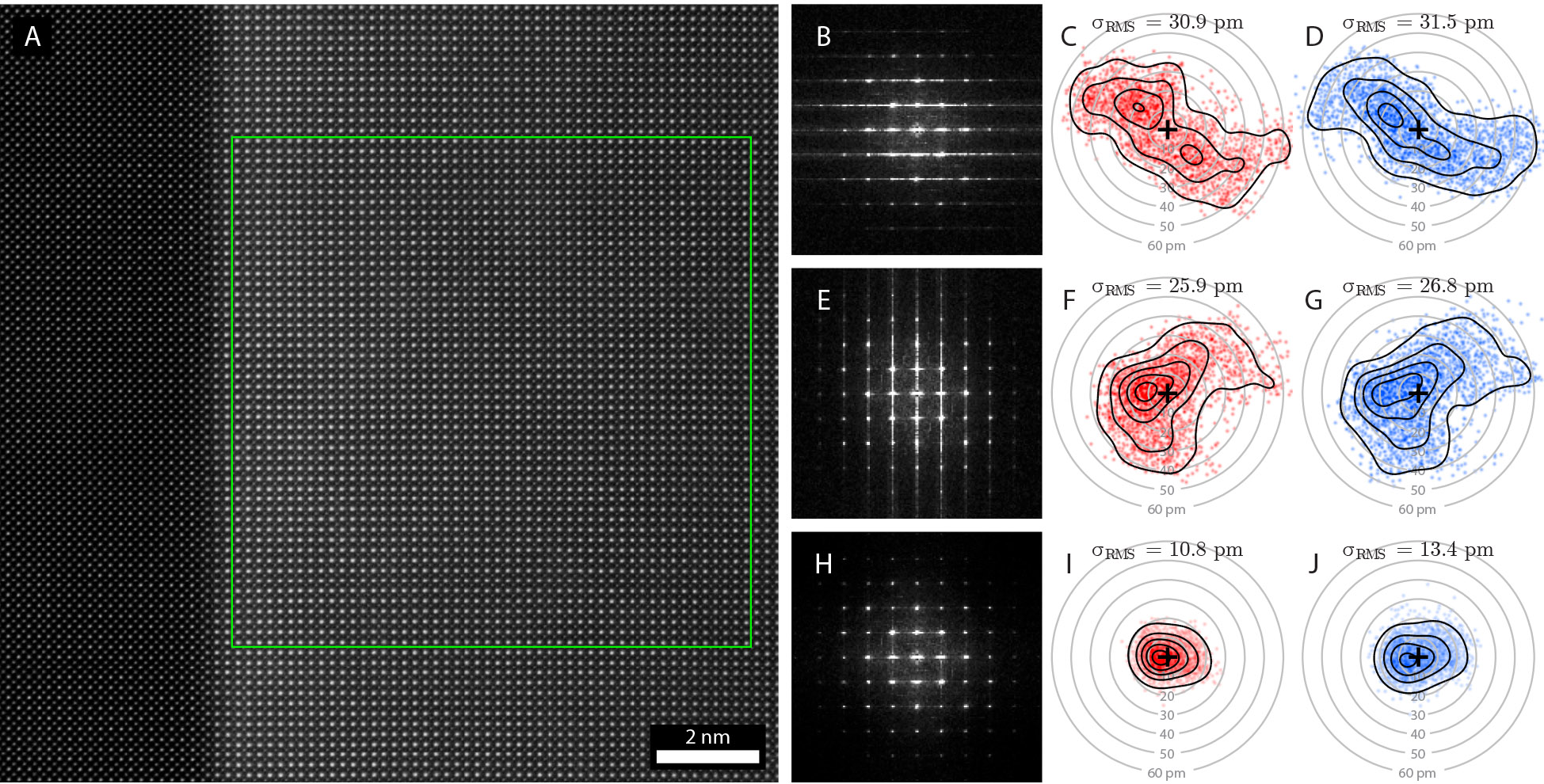}
 	\caption{(A) Drift-corrected HAADF image of a La$_{0.2}$Bi$_{0.8}$FeO$_{3}$ film from an orthogonal scan pair, imaged along the [001] direction. Fourier transform amplitudes of (B) vertical fast scan direction image, (E) horizontal fast scan direction image, and (H) drift-corrected image given in (A).  Two-dimensional deviations from best-fit lattice for box shown in (A) for the La/Bi sites are plotted in (C), (F), and (I), and deviations of the Fe sites in (D), (G), and (J), in the same order as (B), (E) and (H). RMS 2D deviation $\sigma_{\rm{RMS}}$ shown for all sites/images.}
	\label{FigureResultsExpSimple}
\end{figure*}


We next test our algorithm using a simple experimental dataset, where the sample consists of two perovskite layers, a La$_{0.2}$Bi$_{0.8}$FeO$_3$ film on a SrRuO$_3$ buffer electrode, imaged along the [001] zone axis. The detector used creates a high angle annular dark field (HAADF) image that is sensitive to atomic number.  The  La$_{0.2}$Bi$_{0.8}$FeO$_3$  layer appears brighter due to the high-Z elements La $(Z=57)$ and Bi $(Z=83)$ on the perovskite A-site.  The SrRuO$_3$ layer appears dimmer, as its brightest peak occurs for Ru $(Z=44)$ on the perovskite B-site.  Two HAADF images were recorded with orthogonal scan directions, and the resulting corrected image is plotted in Fig.\ \ref{FigureResultsExpSimple}A.  The Fourier transform amplitudes for the original vertical scan image, the original horizontal scan image and the corrected image are plotted in Figs.\ \ref{FigureResultsExpSimple}B, E, and H respectively.  As in the synthetic example of the previous section, the correction routine has removed the streaks along the slow scan directions.  

To quantify the improvement in data quality after reconstruction, we have performed nonlinear least squares fits of 2D Gaussian peaks (5 parameter fits: peak position, peak intensity, standard deviation and a constant background intensity) to all atomic columns in a 50 x 50 unit cell region, shown in Fig.\ \ref{FigureResultsExpSimple}A.  We compared all column position fits to a global best-fit linear lattice, and calculated the deviation $\sigma_{\rm{RMS}}$ from the ideal positions.  A 2D histogram of the La/Bi and Fe sites for the vertical scan image is plotted in Fig.\ \ref{FigureResultsExpSimple}C and D, for the horizontal scan image in Fig.\ \ref{FigureResultsExpSimple}F and G, and for the corrected image in Fig.\ \ref{FigureResultsExpSimple}I and J respectively.  

The results show that for both original images, long-range nonlinear distortions have produced large asymmetric deviations from the ideal lattice.  By comparison, the corrected image deviations form a tighter and more symmetric pattern of deviations from the ideal lattice.  The root-mean-square of all deviations $\sigma_{\rm{RMS}}$ for the La/Bi and Fe sites has been reduced by a factor of approximately three and two respectively.  More importantly, the asymmetric deviations caused by sample drift distortions have been almost completely removed.   



\subsection{Complex experimental example}

\begin{figure*}[htbp]
    \centering
        \includegraphics[width=6.0in]{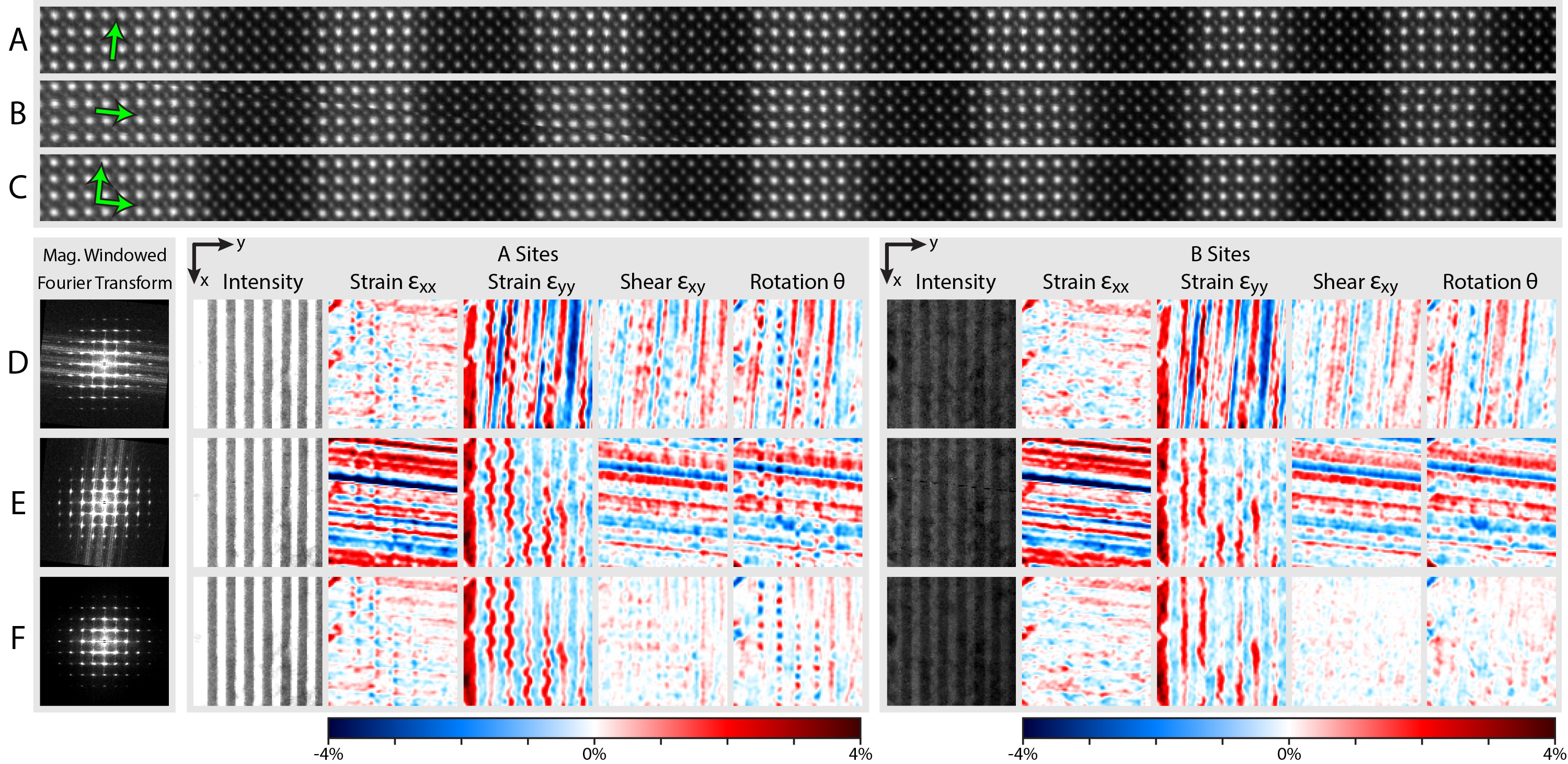}
 	\caption{Example of nonlinear drift correction for a complex experimental dataset.  Strip of a square HAADF image for (A) near-vertical scan direction, (B) near-horizontal scan direction and (C) drift-corrected image from (A) and (B).  (D), (E), and (F) rows contain Fourier transform amplitudes, site intensities and strain field measurements for A and B sites, measured from (A), (B) and (C) respectively.}
	\label{FigureResultsExpComplex}
\end{figure*}

In the previous section, the sample consisted of a nearly ideal lattice, which could have been approximately corrected by assuming an ideal lattice.  To demonstrate that our algorithm works with arbitrarily complex experimental data, we now examine an experimental dataset containing a large amount of nonlinear distortions over many different length scales. This sample is composed of a superlattice of alternating perovskite layers, similar to those described by Schlom et al.\ \cite{schlom2007strain}, with a non-uniform lattice structure as shown in Fig.\ \ref{FigureResultsExpComplex} and detailed in Ref.\ \cite{nelsonSubmission}.  This sample was chosen because these HAADF images contain a large amount of nonlinear drift distortion, and it is known to contain a complex strain field.  Again, we have recorded orthogonal scan pairs and applied our algorithm to correct the drift distortions, using directional Fourier weighting to produce the final image.  Strips from the same region of the original images and the corrected image are plotted in Figs.\ \ref{FigureResultsExpComplex}A, B and C respectively.  A jump discontinuity is visible running across the left side of Fig.\ \ref{FigureResultsExpComplex}B, but it has been removed from Fig.\ \ref{FigureResultsExpComplex}C.  Both of Figs.\ \ref{FigureResultsExpComplex}A and B contain small long-range lattice distortions that are not present in Fig.\ \ref{FigureResultsExpComplex}C. These distortions could easily be mistaken for real variations in the sample without application of nonlinear drift correction.


We have performed a measurement of the lattice strains from both the A (brighter) and the B (dimmer) perovskite sites, for both of the original images and the corrected image.  The lattice strains are calculated using a real space fitting procedure.  First, all site positions are measured using a 2D Gaussian fit.  Then, a best-fit reference lattice was calculated for all sites, and the deviations from the ideal lattice positions are computed.  These deviations are transformed into a 2D image using KDE, as in Fig.\ \ref{FigureKDE}, with a kernel standard deviation of 1 unit cell length.  Finally, the 2D strain maps are calculated by numerical differentiation of the displacement maps.  The site intensities and strain maps for the original images and the corrected image are shown in Figs.\ \ref{FigureResultsExpComplex}D-F.

The strain maps in Figs.\ \ref{FigureResultsExpComplex}D and E show clear signs of drift distortion, manifesting as long streaks along the slow scan directions that span the field of view.  These streaks are especially pronounced in strain measurements, since numerical differentiation magnifies any noise present.  The only strain maps that do not show large streak artifacts are those measured along a direction close to the fast scan direction, $\epsilon_{\rm{xx}}$ for Fig.\ \ref{FigureResultsExpComplex}D, and $\epsilon_{\rm{yy}}$ for Fig.\ \ref{FigureResultsExpComplex}E.  By contrast, no artifacts are visible in the drift-corrected strain maps plotted in Fig.\ \ref{FigureResultsExpComplex}F.  This example demonstrates that our algorithm can successfully correct nonlinear drift without removing complex real signals in the underlying data.  A detailed analysis of these signals is given in Ref.\ \cite{nelsonSubmission}.


\section{Conclusion}

We have described a procedure for correcting nonlinear drift distortion in scanning probe images using orthogonal scan pairs.   Our algorithm corrects drift distortion in each image line-by-line, by moving each scanline origin to minimize the absolute difference between the measured line intensity and the current iteration of the corrected image recorded along an orthogonal direction.  We have successfully applied the algorithm to synthetic data, and both simple and complex experimental datasets.  Our results show that recording only two images is sufficient to remove most of the nonlinear drift error.  Including additional images in the reconstruction will further reduce the error, especially from additional scan directions.  Our algorithm should prove useful for all scanning probe experiments where high precision is required.


\section{Acknowledgements}

We thank Wolfgang Theis, Peter Ercius, Mary Scott and Matt Bowers for helpful discussions.  Work at the Molecular Foundry was supported by the Office of Science, Office of Basic Energy Sciences, of the U.S. Department of Energy under Contract No. DE-AC02-05CH11231.

\section{Appendix}
\subsection{Kernel Density Estimation \label{SectionAppendixKDE}}

\begin{figure}[htbp]
    \centering
        \includegraphics[width=3.00in]{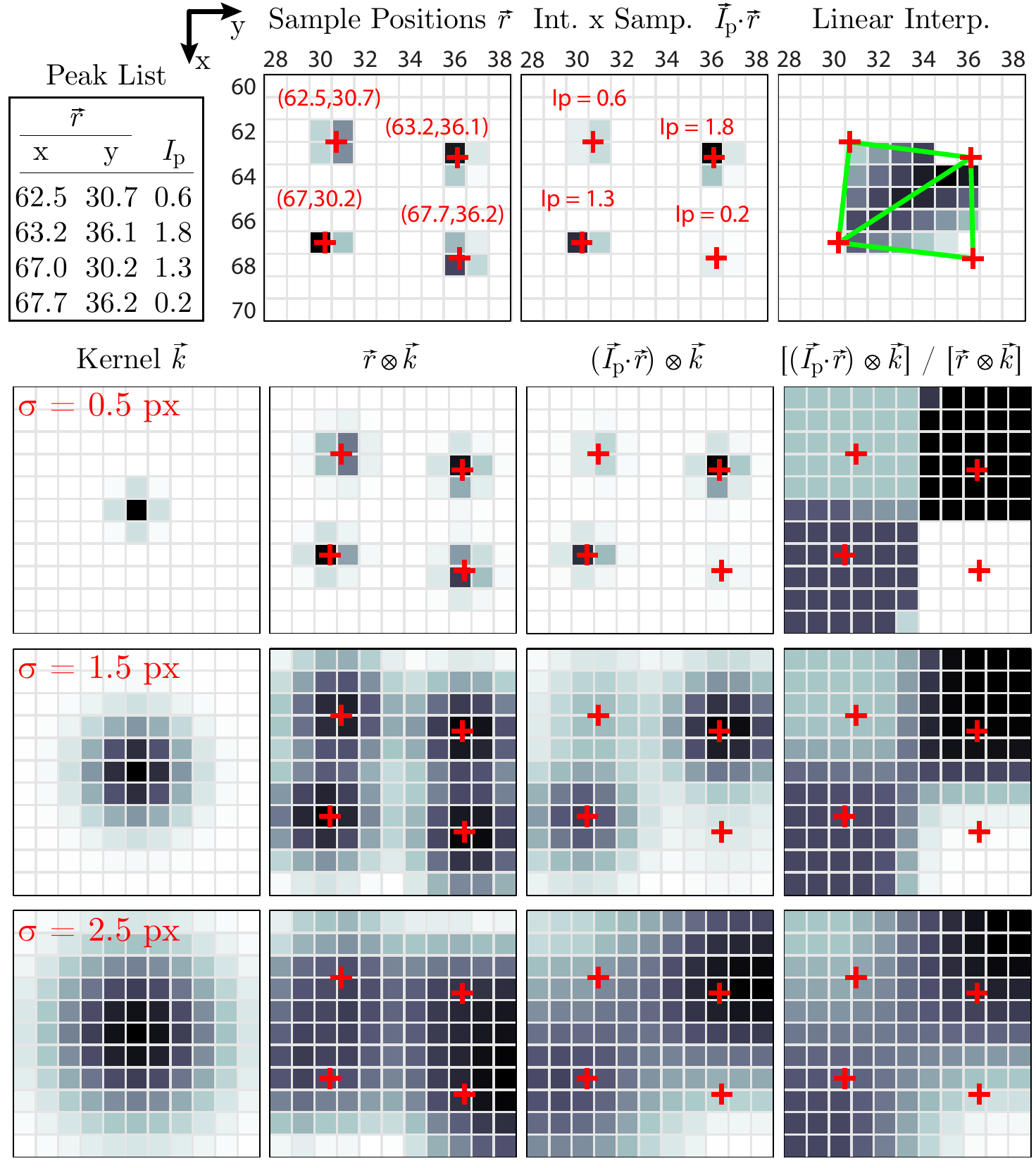}
 	\caption{Top row shows (from left to right) list of peaks with locations $\vec{r}$, bilinear weighting of each sample, bilinear weighting times sample intensity $\vec{I_p}$, and a linear interpolation of these points using triangular regions.  All other rows show grid interpolation of discrete samples using kernel density estimation (KDE), with a different kernel size $\sigma$ for each row.}
	\label{FigureKDE}
\end{figure}

Kernel density estimation (KDE) is a simple and robust method to estimate the values of a continuous field from non-uniformly spaced discrete measurements \cite{rosenblatt1956remarks, parzen1962estimation}.  The procedure we use is demonstrated in Fig.\ \ref{FigureKDE}. All measurement pixels from all scanlines are first added to the reconstruction space using bilinear weighting of a 2x2 pixel region.  Then, KDE is used to interpolate the value of all image pixels.  The ``bandwidth'' of the KDE procedure is set by the standard deviation $\sigma$ of a 2D Gaussian kernel, and is typically set to a value of 0.5 pixels for this algorithm.  If $\sigma$ is much less than the distance between samples, KDE will asymptotically approach a nearest-neighbor interpolation.  As $\sigma$ is increased, the interpolation becomes smoother, shown in Fig.\ \ref{FigureKDE}.  This method is very robust against large nonlinear jumps between adjacent scanlines, where no measurement estimate may be available.  Equally important is the resulting image and its derivatives are free of discontinuities, which is important for peak fitting or strain measurements.  For comparison, a linear interpolation calculated on a nonuniform triangular grid is plotted the upper right plot of Fig.\ \ref{FigureKDE}.  This interpolation produces a poor result with a slope discontinuity at each triangular boundary.

\bibliographystyle{model1-num-names}
\bibliography{STEMrefs}

\end{document}